\documentclass[letterpaper, 10 pt, conference]{ieeeconf}
\IEEEoverridecommandlockouts
\let\proof\relax
\let\endproof\relax
\let\labelindent\relax
\usepackage{enumerate, amsmath, amsthm, amssymb, amsfonts, mathtools, xargs, tensor, units, cite, stmaryrd, mathrsfs, algpseudocode, comment, graphicx}
\usepackage[unicode=true, bookmarks=true, bookmarksnumbered=false, bookmarksopen=false, breaklinks=false, pdfborder={0 0 1}, backref=false, colorlinks=false, hidelinks]{hyperref}

\newcommandx\fid[5][usedefault, addprefix=\global, 1=, 2=, 3=, 4=, 5=]{\tensor*[_{#3}^{#2}]{#1}{_{#5}^{#4}}}

\usepackage[bytype]{Theorems}

\begin{document}
	\title{Switched Optimal Control and Dwell Time Constraints: A Preliminary Study}
	
	\author{Moad Abudia, Michael Harlan, Ryan Self, and Rushikesh Kamalapurkar\thanks{The authors are with the School of Mechanical and Aerospace Engineering, Oklahoma State University, Stillwater, OK, USA. {\tt\small \{abudia, michael.c.harlan, rself, rushikesh.kamalapurkar\}@okstate.edu}.		
	This  research   was  supported,   in  part,  by   the  Air Force  Research  Laboratories  (AFRL)  under  award  number  FA8651-19-2-0009, and the National Science Foundation (NSF) award number 1925147.   Any  opinions,  findings,  conclusions,  or  recommendations  detailed in this article are those of the author(s), and do not necessarily reflect the views of the sponsoring agencies.}}
	\maketitle
	\begin{abstract}
	Most modern control systems are switched, meaning they have continuous as well as discrete decision variables. Switched systems often have constraints called dwell-time constraints (e.g., cycling constraints in a heat pump) on the switching rate. This paper introduces an embedding-based-method to solve optimal control problems that have both discrete and continuous decision variables. Unlike existing methods, the developed technique can heuristically incorporate dwell-time constraints via an auxiliary cost, while also preserving other state and control constrains of the problem. Simulations results for a switched optimal control problem with and without the auxiliary cost showcase the utility of the developed method. 

	\end{abstract}

	\section{Introduction}\label{intro}
	The field of switched optimal control has grown exponentially in recent years due to the advances in computing technology. While optimal control methods for continuous systems have been well-explored, a significant portion of modern controlled systems have continuous as well as binary (or integer) decision variables.  For example, in a refrigeration system with multiple fixed speed compressors and multiple cooling racks, the compressors and the valves represent binary on/off decision variables, yet the dynamics of the system are continuous. Another example is the automatic transmission in a vehicle, which has a discrete number of gears but continuous dynamics for the torque from the engine and load from the wheels. Due to the presence of discrete decision variables, switched optimal control problems (SOCPs) cannot be solved using traditional gradient-based methods developed for continuous optimal control problems (OCPs). Direct transcription of OCPs with continuous and discrete decision variables often results in mixed-integer nonlinear programs that are computationally intensive to solve.

A variety of methods for solving SOCPs have been developed over the past several decades, which can be roughly classified as follows. In the first category are methods that fix a mode sequence \cite{SCC.Xu.Antsaklis2004,SCC.Kamgarpour.Tomlin2012}, or use predefined switching surfaces \cite{SCC.Dharmatti.Ramaswamy2005,SCC.Axelsson.Boccadaro.ea2008} and optimize the timing of switching using smooth optimization techniques such as gradient descent. 

In the second category are methods to solve SOCPs that optimize both the mode sequence and the switching time instances \cite{SCC.Dharmatti.Ramaswamy2005,SCC.Lu.Ferrari2013,SCC.Hedlund.Rantzer1999}.
 In \cite{SCC.Azhmyakov2012} a theoretical framework is developed for a general OCP involving a sub-class of switched systems. In \cite{SCC.Fahroo.Ross2008} a method is developed based on approximations of the differential constraints that are assumed to be given in the form of controlled differential inclusions. However, techniques in the first and second categories do not incorporate dwell-time constraints, and as such are not applicable to the problem at hand.

In the third category, OCPs for switched systems with dwell-time are analyzed using a dynamic programming method for restricted classes of systems such as systems having stable state matrices for all of the subsystems and for all of the time instances\cite{SCC.Jungers.Daafouz2013}, and discrete time autonomous systems \cite{SCC.Heydari2017}. Mode insertion methods such as \cite{SCC.Wardi.Egerstedt.ea2015} can incorporate dwell-time constraints, however, the dwell-time is implemented by filtering the optimal controller after-the-fact, resulting in loss of satisfaction guarantees for other state and control constraints of the problem. The embedding method in \cite{SCC.Bengea.DeCarlo2005}, when applied to problems that have chattering solution, also requires changes to the optimal mode sequence after-the-fact, which results in the loss of satisfaction guarantees for the boundary constraints of the problem.

The goal in this paper is to develop a method to optimize continuous controllers, mode sequences, and mode switching times, while guaranteeing satisfaction of boundary constraints and dwell-time constraints. The paper extends the embedding approach introduced by \cite{SCC.Bengea.DeCarlo2005}, which involves solving the SOCP as a continuous OCP, to heuristically incorporate dwell-time constraints and to preserve boundary constraints by avoiding after-the-fact modification of the optimal mode sequence. The switched system is embedded in a continuous system by implementing the switched control signals as continuous variables, as explained in Section \ref{embedded}. Section \ref{modified} details the primary contributions of this paper, which is adding an auxiliary cost to the cost function in the problem formulation to force a bang-bang type solution. Section \ref{simulation 1} presents simulation results for a simple nonlinear example and discusses how the magnitude of the added auxiliary cost heuristically controls the switching rate of the binary/integer control variables in numerical solutions. Section \ref{Analysis} analyzes the numerical method used in section \ref{simulation 1} and explains the differences between the theoretical and simulation results. Section \ref{conclusion} concludes the effort with a discussion about the ways the developed technique improves the current embedding methods. 

	\section{Embedded Optimal Control Formulation}\label{embedded}
For ease of exposition, this paper focuses on a SOCP with two subsystems, but the developed method can be extended to switched systems with a higher number of subsystems using techniques similar to \cite{SCC.Bengea.DeCarlo2005}. The system state is represented by  $x:\mathbb{R}_{\ge 0}\rightarrow\mathbb{R}^{n}$ with dynamics\footnote{$\mathbb{R}_{\ge a}$ represents all positive real numbers greater than or equal to $a$, and the symbol $\forall\forall$ is used as as a shorthand for the phrase 'for almost all'.}
\begin{align}
\dot{x}(t)=f_{v(t)}(t,x(t),u(t)), \forall\forall t\in[t_0,t_f],
\label{eq:switchedstate}&
\end{align}
and $x(t_0)=x_0$, where $v:\mathbb{R}_{\ge 0}\rightarrow\{0,1\}$ is the mode sequence, $u:\mathbb{R}_{\ge 0}\rightarrow\Omega\subset\mathbb{R}^{m}$ is the control input constrained to the compact set $\Omega$, and $f_{v(t)}\in\mathcal{C}^{1}(\mathbb{R}\times\mathbb{R}^{n}\times\mathbb{R}^{m},\mathbb{R}^{n}),\:\forall t\in[t_0,t_f]$. The control functions $t\mapsto v(t)$ and $t\mapsto u(t)$ must be selected such that the following constraints $(t_0,x(t_0))\in\mathcal{T}_0\times\mathcal{B}_0$ and $(t_f,x(t_f))\in\mathcal{T}_f\times\mathcal{B}_f$ are satisfied, where the endpoint constraint set $\mathcal{B}\coloneqq\mathcal{T}_0\times\mathcal{B}_0\times\mathcal{T}_f\times\mathcal{B}_f\subseteq\mathbb{R}^{2n+2}$ is compact. The switched cost functional is defined as
\begin{multline}
J(t_0,x_0,u\left(\cdot\right),v\left(\cdot\right))\coloneqq\int_{t_0}^{t_f}L_{v(t)}(t,x(t),u(t))\mathrm{d}t\\+K(t_0,x_0,t_f,x_f)\label{eq:switchedcost}
\end{multline}
where $L_{v(t)}\in\mathcal{C}^{1}(\mathbb{R}\times\mathbb{R}^{n}\times\mathbb{R}^{m},\mathbb{R})\:\forall t\in[t_0,t_f]$, and $K\in\mathcal{C}^{1}(\mathbb{R}\times\mathbb{R}^{n}\times\mathbb{R}\times\mathbb{R}^{n},\mathbb{R})$. The SOCP is formulated as

\begin{align*}
	\min_{u\left(\cdot\right),v\left(\cdot\right)} \quad & J(t_0,x_0,u\left(\cdot\right),v\left(\cdot\right)) \quad \textnormal{subject to:}\\
	& \textnormal{(i) } x\left(\cdot\right) \textnormal{satisfies } \eqref{eq:switchedstate}, \\
	& \textnormal{(ii) } (t_0,x(t_0),t_f,x(t_f))\in\mathcal{B}, \\
	& \textnormal{(iii) } v(t)\in\{0,1\}, u(t)\in\Omega,\:\forall t\in[t_0,t_f],\\
	& \textnormal{(iv) } \forall t_1 , t_2 \in [t_0, t_f] \textnormal{ with } v(t_1^-)\ne v(t_1^+)\textnormal{ and }\\ 
	&  \quad \quad  v(t_2^-)\ne v(t_2^+), |t_1-t_2|\geq T>0,
\end{align*}
where (iv) encodes the dwell-time constraint. In order to solve this problem using conventional gradient-based techniques that stem from dynamic programming or Pontryagin's minimum principle, all the decision variables in the optimization problem need to be continuous. Therefore the SOCP is embedded in a larger domain by replacing the mode sequence with a continuous function $\bar{v}:\mathbb{R}_{\ge 0}\rightarrow[0,1]$. Letting $u_i:\mathbb{R}_{\ge 0}\rightarrow\Omega, \forall i\in\{0,1\}$ be the control input in vector field $f_i$, the dynamics of the embedded system are defined as
\begin{multline}   
\dot{x}(t)\coloneqq[1-\bar{v}(t)]f_0(t,x(t),u_0(t))+\bar{v}(t)f_1(t,x(t),u_1(t)),\\x(t_0)=x_0\:,\forall\forall t\in[t_0,t_f],\label{eq:embeddedstate}
\end{multline}
and the cost functional is defined as
\begin{multline}
J(t_0,x_0,u_0\left(\cdot\right),u_1\left(\cdot\right),\bar{v}\left(\cdot\right))\coloneqq\\\int_{t_0}^{t_f}\Big( [1-\bar{v}(t)]L_0(t,x(t),u_0(t))\\+\bar{v}(t)L_1(t,x(t),u_1(t))\Big)\mathrm{d}t+K(t_0,x_0,t_f,x_f)\label{eq:embeddedcost}
\end{multline}
The embedded optimal control problem (EOCP) is then formulated as 
\begin{align}
\min_{u_0\left(\cdot\right),u_1\left(\cdot\right),\bar{v}\left(\cdot\right)} \quad & J\left(t_0,x_0,u_0\left(\cdot\right),u_1\left(\cdot\right),\bar{v}\left(\cdot\right)\right) \quad \textnormal{subject to:} 
\label{eq:embeddedproblem}
\end{align}
\begin{align*}
& \textnormal{(i) } x\left(\cdot\right) \textnormal{ satisfies } \eqref{eq:embeddedstate}, \\
& \textnormal{(ii) } (t_0,x(t_0),t_f,x(t_f))\in\mathcal{B},\\
& \textnormal{(iii) } \bar{v}(t)\in[0,1],\:u_0(t),u_1(t)\in\Omega,\:\forall t\in[t_0,t_f].
\end{align*}
Since $\bar{v}\left(\cdot\right)$ is continuous, all the decision variables of the EOCP are continuous, and as such, the classical necessary and sufficient conditions of optimal control theory can be utilized to analyze the EOCP \cite{SCC.Berkovitz1974}. However, solutions of the EOCP are not necessarily feasible for the SOCP, since the EOCP allows for smooth transition between modes, while the SOCP does not. Furthermore, the dwell-time constraint (iv) of the SOCP cannot be easily incorporated in the ECOP. To ensure that the EOCP produces bang-bang solutions that are feasible for the SOCP, the embedding process is modified in Section III.

	\section{Modified Formulation}\label{modified}
When a solution to the EOCP is a regular solution, otherwise known as a bang-bang type solution, the range of the embedded mode signal, $\bar{v}$, is restricted to $\{0,1\}$, so it can be identified with a switched mode sequence, $v$. In fact, regular solutions of the EOCP are also solutions of the SOCP, as shown in \cite{SCC.Bengea.DeCarlo2005}. However, whenever the EOCP has a singular solution, i.e., $\bar{v}(t)\in(0,1)$ over a time interval of nonzero measure, the embedded mode signal cannot be identified directly with any switched mode sequence. Using the Chattering Lemma \cite{SCC.Ge1996}, a sub-optimal bang-bang solution can be determined within an arbitrary $\epsilon$-distance from the singular solution \cite{SCC.Bengea.DeCarlo2005}. However, such a solution may not satisfy the boundary constraints and the path constraints of the SOCP.

 The method developed in this paper ensures that the EOCP produces bang-bang solutions that meet all the boundary and path constraints, while heuristically accounting for dwell-time constraints of the SCOP.

Motivated by the Hamiltonian minimization condition of Pontryagin's minimum principle, a concave auxiliary cost function $L_{\bar{v}}:[0,1]\rightarrow\mathbb{R}$, such that $L_{\bar{v}}\left(\bar{v}(t)\right)=0$ whenever $\bar{v}\left(t \right)\in\{0,1\}$ , and $L_{\bar{v}}\left(\bar{v}(t)\right)>0$ whenever $\bar{v}\left(t\right)\in\left(0,1\right)$ is added to the EOCP to enforce bang-bang solutions.  For example, an inverted parabola of the form $L_{\bar{v}}(\bar{v}\left(\cdot\right))=4\beta(\bar{v}\left(\cdot\right)-\bar{v}\left(\cdot\right)^2)$ with $\beta$ being a positive constant could be used, which outputs 0 at $\bar{v}\left(t\right)\in\{0,1\}, \forall t\in[t_0, t_f]$ and reaches a maximum value of $\beta$ when $\bar{v}\left(\cdot\right)\equiv0.5$. Since the minima of the auxiliary cost function will be the modes of the SOCP, in minimizing the cost functional to find an optimal solution to the modified ECOP (MEOCP), the embedded mode signal is pushed towards the modes of the SOCP and away from intermediate states which are not feasible for the SOCP. In the following, Pontryagin's minimum principle is used to prove that the solutions of the MEOCP are of a bang-bang type, which can be directly implemented on the switched system, allowing the SOCP to be solved as a continuous OCP using conventional techniques.\footnote{Design of auxiliary cost functions for a switched system with three or more subsystems, where the Hamiltonian is equal to zero when evaluated on the vertices of a polygon, is part of future research.}

Let $V:=[u_0,u_1,\bar{v}]^T$ denote the augmented control input. The Hamiltonian for the MEOCP is defined as 
\begin{multline} 
H(x,\lambda,V,t)\coloneqq\langle\lambda,[1-\bar{v}]f_0(t,x,u_0)+\bar{v}f_1(t,x,u_1)\rangle\\
+[1-\bar{v}]L_0(t,x,u_0)+\bar{v}L_1(t,x,u_1)+L_{\bar{v}}(\bar{v}).\label{eq:hamiltonian}
\end{multline}
According to Pontryagin's minimum principle, when evaluated along the optimal state trajectory, $x^*\left(\cdot\right)$, and the optimal costate trajectory, $\lambda^*\left(\cdot\right)$, the optimal augmented control $V^*\left(\cdot\right)=[u^*_0\left(\cdot\right),u^*_1\left(\cdot\right),\bar{v}^*\left(\cdot\right)]^T$ minimizes the Hamiltonian among all other controllers. That is, $\forall t$, $V^*(t)\in\mathbb{R}^3$ minimizes the function $V\mapsto H(x^*(t),\lambda^*(t),V,t)$.

As a result, it can be concluded that the function $\bar{v}\mapsto H(x^*(t),\lambda^*(t),[u^*_0(t),u^*_1(t),\bar{v}]^T,t)$ is minimized by the optimal mode signal $\bar{v}^*(t),\:\forall t$. Since $L_{\bar{v}}$ is a concave function and since the function $\bar{v}\mapsto \langle\lambda^*(t),[1-\bar{v}]f_0(t,x^*(t),u^*_0(t)) +\bar{v}f_1(t,x^*(t),u^*_1(t))\rangle +[1-\bar{v}]L_0(t,x^*(t),u^*_0(t)+\bar{v}L_1(t,x^*(t),u^*_1(t))$ is affine $\forall t$, the function $\bar{v}\mapsto H(x^*(t),\lambda^*(t),[u^*_0(t),u^*_1(t),\bar{v}]^T,t)$ is a sum of a concave function and an affine function, and as a result, is concave for all $t$. Since concave functions over a compact set have their minima on the boundary of the compact set (see \cite[Theorem 3]{SCC.Zangwil1967}), the function $\bar{v}\mapsto H(x^*(t),\lambda^*(t),[u^*_0(t),u^*_1(t),\bar{v}]^T,t)$ is minimized at the boundary of the range of the embedded mode signal.

In the two-switched system, the Hamiltonian is minimized when the embedded mode signal is either 0 or 1, which can be identified with the modes of the SOCP. Thus, the solution found for the MEOCP can be implemented in the original switched system. In summary
 the auxiliary cost ensures that solutions of the MEOCP are of a bang-bang type, and as a result, can be implemented on the original switched system. However, bang-bang optimal solutions of the MEOCP, while feasible, are not necessarily optimal for SOCP. Determination of optimality of bang-bang solutions of the MEOCP with respect to the SOCP is an open question that requires further research. 

	\section{Simulation Example}\label{simulation 1}
An example is used to showcase the developed method, featuring a valve which empties into a tank which empties into a second tank and then out. The flow rate out of each tank is modeled as the square root of the water level, which is the state of the system, and the objective is to maintain a given water level in the second tank, with the control input being the switched flow rate of the input valve. The valve can either be open at a high flow rate of 2 or at a low flow rate of 1. The dynamics of this switched system are defined as
\begin{equation}
\begin{bmatrix}
\dot{x}_1(t)\\\dot{x}_2(t)
\end{bmatrix}=
\begin{cases}
\begin{bmatrix}
1-\sqrt{x_1(t)}\\\sqrt{x_1(t)}-\sqrt{x_2(t)}
\end{bmatrix},
& v(t)=0,\\

&   \\
\begin{bmatrix}
2-\sqrt{x_1(t)}\\\sqrt{x_1(t)}-\sqrt{x_2(t)}
\end{bmatrix},
& v(t)=1,
\end{cases}\label{eq:switchedexmp}
\end{equation}
$\forall\forall t\in[t_0,t_f]$, where $x_1\left(\cdot\right)$ is the water level in the first tank, $x_2\left(\cdot\right)$ is the water level in the second tank, and $v\left(t\right)\in\{0,1\}, \forall t \in [t_0,t_f]$, denotes the mode of operation. 
The cost functional to be minimized is defined as
\begin{align}
J(t_0,x_0,v\left(\cdot\right))=\int_{t_0}^{t_f}\alpha(x_2(t)-3)^{2}\mathrm{d}t.
\end{align}
with $t_0=0$, $t_f=20$, $\alpha=2$, and $x(0)=\begin{bmatrix}2 \ 2\end{bmatrix}^T$. The goal is to achieve a water level of 3 in the second tank, which yields the SOCP  
\begin{align*}
\min_{v(\cdot)} \quad & J(t_0,x_0,v\left(\cdot\right)) \quad \textnormal{subject to:}\\
& \textnormal{(i) } x\left(\cdot\right) \textnormal{ satisfies } \eqref{eq:switchedexmp}, \\
& \textnormal{(ii) } (t_0,x(t_0),t_f,x(t_f))\in\\
&\quad\quad\{0\}\times\begin{bmatrix}2\\2\end{bmatrix}\times\{20\}\times\begin{bmatrix}[0,4]\\3\end{bmatrix}, \\
& \textnormal{(iii) } v(t)\in\{0,1\},\:\forall t\in[t_0,t_f].
\end{align*}
For the water level in the second tank to be maintained at 3,  we need $\dot{x}_2(t)=0$, which means $\sqrt{x_1(t)}=\sqrt{x_2(t)}\implies x_1(t)=x_2(t)=3$. This means the water level in the first tank must also be 3, and as a result, $\dot{x}_1(t)=0$. Therefore, the flow rate of the valve must be $\sqrt{x_1(t)}=\sqrt{3}$ which is between the high and the low valve flow rates. A flow rate of $\sqrt{3}$ is impossible to achieve without the solution infinitely fast switching between the high and the low valve state, which is characteristic of a singular solution to the SOCP \cite{SCC.Bengea.DeCarlo2005}.

 Following the developed embedding method, the constraints on the control are modified by the introducing the embedded mode signal $\bar{v}:[t_0,t_f]\to[0,1]$. The dynamics then can be formulated using (\ref{eq:embeddedstate}) as
\begin{multline}
\begin{bmatrix}
\dot{x}_1(t)\\\dot{x}_2(t)
\end{bmatrix}=
\begin{bmatrix}
1+\bar{v}(t)-\sqrt{x_1(t)}\\\sqrt{x_1(t)}-\sqrt{x_2(t)}
\end{bmatrix},
x(t_0)=\begin{bmatrix}2\\2\end{bmatrix},\:\\
\forall\forall t\in[t_0,t_f].
\label{eq:embeddedexmp}
\end{multline}
Using the cost functional in (\ref{eq:switchedcost}), the EOCP is given by (\ref{eq:embeddedproblem}). 

\begin{figure}
\includegraphics[width=\linewidth]{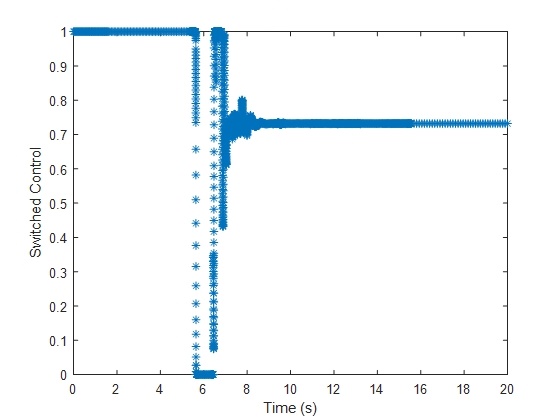}
\caption{The optimal control signal for the unmodified two-tank system EOCP.}
\label{fig:controlwocost}
\end{figure}

\begin{figure}
\includegraphics[width=\linewidth]{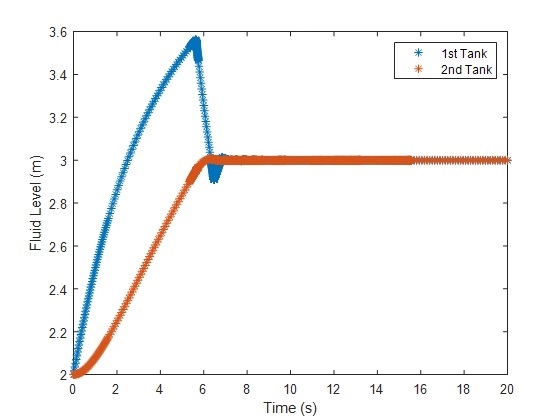}
\caption{The states of the two-tank system under optimal control signal for the unmodified EOCP, with a final cost of 4.7312.}
\label{fig:statewocost}
\end{figure}

The results in Fig \ref{fig:controlwocost} indicate that the solution of the EOCP is singuler, and as such, cannot be directly implemented in the original switched system. The ECOP is solved using pseudospectral optimization via GPOPS \cite{SCC.Rao.Benson.ea2010}, with the differentiation method set to finite-difference and the maximum mesh iterations set to 5. As can be seen in Fig. \ref{fig:controlwocost}, the control signal settles to a value of $\approx\sqrt{3}$, in agreement with the analytic solution. As shown in Fig. \ref{fig:statewocost} both of the tank levels settle to the desired level, 3.

Motivated by the discussion in section \ref{modified}, an auxiliary cost function of the form $4\beta(\bar{v}\left(\cdot\right)-\bar{v}^{2}\left(\cdot\right))$ is added to the EOCP to formulate the MEOCP. The auxiliary cost is zero at both extremes of the embedded mode signal and reaches its maximum, $\beta$, in between. The  auxiliary cost function was derived using the embedding process in section \ref{modified}  by setting $L_0(t,x(t),\bar{v}(t))=2\beta \bar{v}(t)+\alpha(x_2(t)-3)^2$ and $L_1(t,x(t),\bar{v}(t))=2\beta(1-\bar{v}(t))+\alpha(x_2(t)-3)^2$, which after simplifying, yields the cost functional for the MEOCP as
$J(t_0,x_0,\bar{v}\left(\cdot\right)=\int_{0}^{20}(\alpha(x_2(t)-3)^2+4\beta(\bar{v}(t)-\bar{v}^2(t))\mathrm{d}t.$
\begin{figure}
\includegraphics[width=\linewidth]{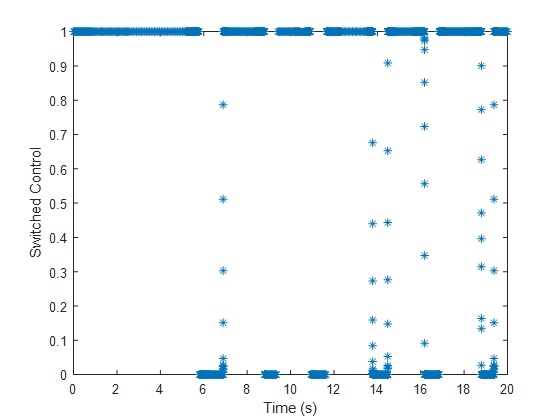}
\caption{The optimal control signal for the two-tank system EOCP with $\beta=0.01$.}
\label{fig:controlw001}
\end{figure}
\begin{figure}
\includegraphics[width=\linewidth]{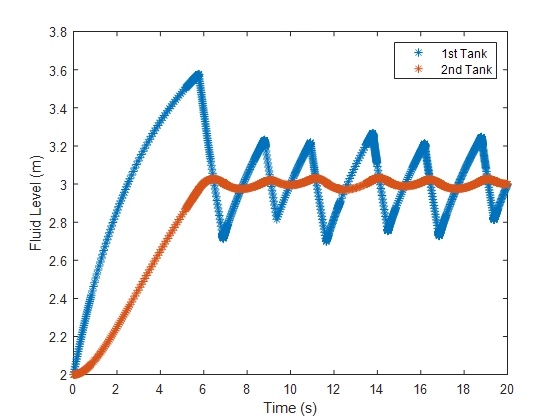}
\caption{The states of the two-tank system under the optimal control signal for the EOCP, with $\beta=0.01$, with a final cost of 4.7355.}
\label{fig:statew001}
\end{figure}
\begin{figure}
	\includegraphics[width=\linewidth]{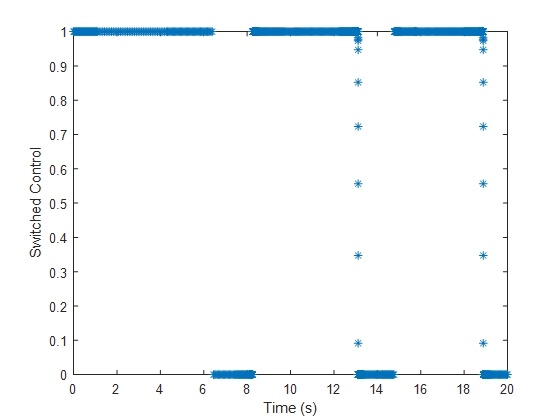}
	\caption{The optimal control signal for the two-tank system EOCP with $\beta=0.2.$}
	\label{fig:controlw02}
\end{figure}

\begin{figure}
	\includegraphics[width=\linewidth]{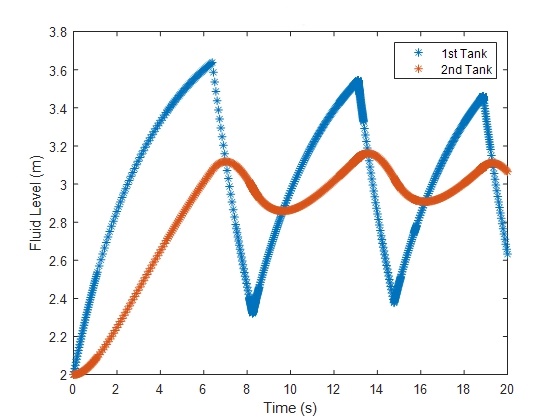}
	\caption{The states of the two-tank system under the optimal control signal for the EOCP, with $\beta=0.2$, with a final cost of 4.8032.}
	\label{fig:statew02}
\end{figure}

The numerical results indicate that the magnitude of the added switching cost can be modified in accordance with the desired dwell-time constraint, with a higher magnitude resulting in a longer dwell-time, and vice versa. Figs. \ref{fig:controlw001} and \ref{fig:statew001} show the solution to the modified EOCP when $\beta=0.01$, which results in a small auxiliary cost relative to the total cost, a linear switching rate, and a lower dwell-time.\\
In each solution run with a different value of $\beta$, the control is of a bang-bang type, which can be directly implemented in the original switched system. Changing $\beta$ changes the frequency of the switching, thus a dwell-time constraint can be implemented heuristically by tuning $\beta$. A greater value of $\beta$ results in a lower switching rate, but the water level in the second tank is not kept as close to the desired level, and the optimal cost is higher.
Figs. \ref{fig:controlw001} and \ref{fig:controlw02} indicate that increasing the constant $\beta$ in the auxiliary cost function increases the time between two consecutive switches of the solution generated by the EOCP solver. 

Since the auxiliary cost function evaluates to zero when the embedded mode
sequence takes the values 0 or 1, the contribution of this function to the total cost should be
zero when the solution of the EOCP is bang-bang. Furthermore, the costate dynamics are
also independent of the auxiliary cost function. \textit{It can be shown that there should not
be any correlation between $\beta$ and the dwell-time of the bang-bang solution and that
Pontryagin’s minimum principle fails to explain the correlation observed in the numerical results}. In fact, a heuristic
examination of the two-tank MEOCP reveals that the optimal solution is a sliding mode
solution that keeps the water level in the second tank at exactly the required height using
infinite frequency switching. 


%
 	\section{Analysis of the Numerical Results}\label{Analysis}
While the auxiliary cost, evaluated along any trajectory where $\bar{v}(t) \in \{0,1\}$, $\forall \in [t_0, t_f]$, is identically zero, solutions of the MEOCP, computed using numerical methods that rely on smoothness of the state and control trajectories, generally include intervals of time where $\bar{v}(t) \notin \{0,1\}$. As a result, it is hypothesized that while an optimal bang bang solution of the MEOCP is independent of the auxiliary cost, the numerical approximation of the optimal solution does depend on the auxiliary cost, and as a result, the parameter $\beta$. This section presents numerical experiments to support the above hypothesis

According to \cite{SCC.Rao.Benson.ea2010}, GPOPS uses Legendre–Gauss–Radau points to determine where to do orthogonal collocation. Collocation is an interpolation technique on continuous functions. As a result, when calculating the cost, the solver interpolates between the high and the low embedded mode signal values in the collocation points on either side of a mode switching time instance. As a result, for a few collocation points, the embedded mode signal is no longer at 0 or 1, and at those collocation points, the trajectory incurs a cost dependent on $\beta$. 

\begin{figure}
	\includegraphics[width=\linewidth]{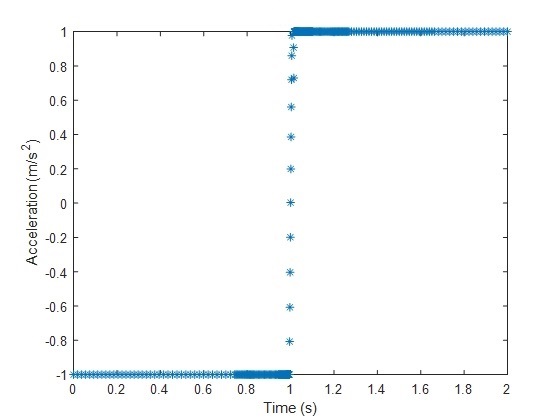}
	\caption{The optimal control signal computed using GPOPS for the double integrator system, with $\beta=0$.}
	\label{fig:vehiclecontrol}
\end{figure}
\begin{figure}
	\includegraphics[width=\linewidth]{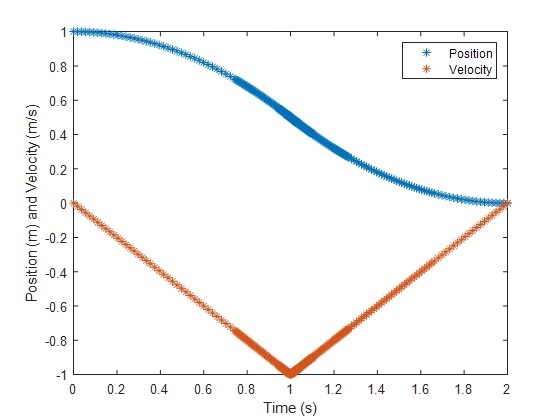}
	\caption{The optimal states trajectories computed using GPOPS for the double integrator system, with $\beta=0$.}
	\label{fig:vehiclestate}
\end{figure}
\begin{figure}
	\includegraphics[width=\linewidth]{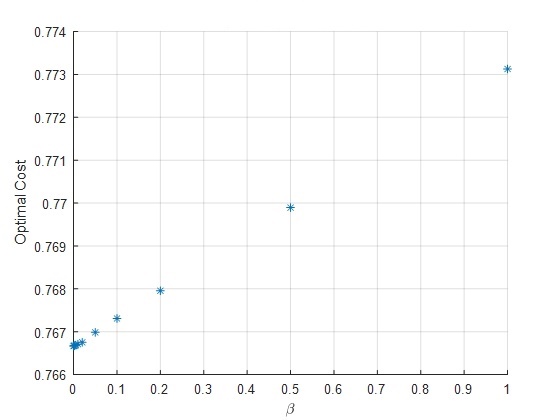}
	\caption{The relationship between $\beta$ and the optimal cost calculated by the numerical solver for the double integrator system.}
	\label{fig:betaVcost}
\end{figure}

The hypothesis can be tested using the following problem with a known solution, as shown in Fig. \ref{fig:vehiclecontrol} and \ref{fig:vehiclestate}. Consider a double integrator 
\begin{equation}
\begin{bmatrix}
\dot{x}_1(t)\\\dot{x}_2(t)
\end{bmatrix}=
\begin{bmatrix}
x_2(t)\\u(t)
\end{bmatrix}, \quad
x(t_0)=\begin{bmatrix}1\\0\end{bmatrix}, \quad x(t_f)=\begin{bmatrix}0\\0\end{bmatrix},
\label{eq:vehicle}
\end{equation}
 $\forall\forall t\in[0,2]$, where the control signal is constrained as $u\left(t\right)\in[-1,1], \forall t \in [0,2]$. The cost function for this system is defined as
\begin{align}
J(t_0,x_0,u\left(\cdot\right))=\int_{t_0}^{t_f}x_1^2(t)+\beta(1-u^2(t))\mathrm{d}t.
\end{align}
This problem can be solved analytically for $\beta=0$, with the optimal control being 
\begin{equation}
u(t)=
\begin{cases}

-1, \quad \forall t\in[0,1),

&   \\
1, \quad \forall t\in[1,2],

\end{cases}\label{eq:switchedexmp}
\end{equation} and the minimal cost being 23/30. 

Since the contribution of the auxiliary cost, $\beta(1-v^2(t))$, is zero along the optimal trajectory, the optimal cost, for $\beta \neq 0$, should be independent of $\beta$. However, as indicated by Fig. \ref{fig:betaVcost}, the value of $\beta$ changes the optimal cost, when computed numerically.

Psedospectral numerical optimal control methods can be used in multi-phase modes, where a mesh division occurs at a switching point, and each section of the trajectory is calculated separately, avoiding interpolation over the switching point. As a result, the cost, computed using a multi-phasic implementation, would no longer depend on $\beta$.
However, such a mesh division is possible only when the time of switching is known a priori, which is not the case in the SOCP studied in this paper.

	\section{Conclusion}\label{conclusion}
This paper introduces a way to solve SOCPs with dwell-time constraints via embedding, by imposing a cost on the embedded mode signal to encourage solutions of a bang-bang type, which can be directly identified with modes switched system. A dwell-time constraint can be implemented heuristically using this method by changing the magnitude of the constant introduced in the cost on the embedded mode signals. Unlike existing methods, in spite of the heuristics involved in tuning the dwell-time, the solutions produced are guaranteed to meet other state and control constraints of the original problem. The embedding facilitates the use of fast, gradient-based numerical methods to solve the SOCP.

To capture the relationship between $\beta$ and the optimal trajectory theoretically, a constrained version of the MEOCP can be formulated, where the embedded mode signal is constrained  to be continuous with respect to time. Formulation and analysis of the constrained version is a topic for future research.

\bibliographystyle{IEEEtran}
\bibliography{sccmaster,scc}

\begin{thebibliography}{10}
\def\url#1{}
\csname url@samestyle\endcsname
\providecommand{\newblock}{\relax}
\providecommand{\bibinfo}[2]{#2}
\providecommand{\BIBentrySTDinterwordspacing}{\spaceskip=0pt\relax}
\providecommand{\BIBentryALTinterwordstretchfactor}{4}
\providecommand{\BIBentryALTinterwordspacing}{\spaceskip=\fontdimen2\font plus
\BIBentryALTinterwordstretchfactor\fontdimen3\font minus
  \fontdimen4\font\relax}
\providecommand{\BIBforeignlanguage}[2]{{%
\expandafter\ifx\csname l@#1\endcsname\relax
\typeout{** WARNING: IEEEtran.bst: No hyphenation pattern has been}%
\typeout{** loaded for the language `#1'. Using the pattern for}%
\typeout{** the default language instead.}%
\else
\language=\csname l@#1\endcsname
\fi
#2}}
\providecommand{\BIBdecl}{\relax}
\BIBdecl

\bibitem{SCC.Xu.Antsaklis2004}
X.~Xu and P.~J. Antsaklis, ``Optimal control of switched systems based on
  parameterization of the switching instants,'' \emph{IEEE Trans. Autom.
  Control}, vol.~49, no.~1, pp. 2--16, 2004.

\bibitem{SCC.Kamgarpour.Tomlin2012}
M.~Kamgarpour and C.~Tomlin, ``On optimal control of non-autonomous switched
  systems with a fixed mode sequence,'' \emph{Automatica}, pp. 1177--1181,
  2012.

\bibitem{SCC.Dharmatti.Ramaswamy2005}
S.~Dharmatti and M.~Ramaswamy, ``Hybrid control systems and viscosity
  solutions,'' \emph{SIAM J. Control Optim.}, pp. 1259--1288, 2005.

\bibitem{SCC.Axelsson.Boccadaro.ea2008}
H.~Axelsson, M.~Boccadaro, M.~Egerstedt, P.~Valigi, and Y.~Wardi, ``Optimal
  mode-switching for hybrid systems with varying initial states,''
  \emph{Nonlinear Anal. Hybrid Syst.}, pp. 765--772, 2008.

\bibitem{SCC.Lu.Ferrari2013}
W.~Lu and S.~Ferrari, ``An approximate dynamic programming approach for
  model-free control of switched systems,'' in \emph{Proc. IEEE Conf. Decis.
  Control}, 2013, pp. 3837--3844.

\bibitem{SCC.Hedlund.Rantzer1999}
S.~Hedlund and A.~Rantzer, ``Optimal control of hybrid systems,'' in
  \emph{Proc. IEEE Conf. Decis. Control}, vol.~4, 1999, pp. 3972--3977.

\bibitem{SCC.Azhmyakov2012}
V.~Azhmyakov, ``On the set-valued approach to optimal control of sliding mode
  processes,'' \emph{J. Franklin Inst.}, vol. 349, no.~4, pp. 1323 -- 1336,
  2012.

\bibitem{SCC.Fahroo.Ross2008}
F.~Fahroo and I.~M. Ross, ``Pseudospectral methods for infinite-horizon
  nonlinear optimal control problems,'' \emph{J. Guid. Control Dynam.},
  vol.~31, no.~4, pp. 927--936, 2008.

\bibitem{SCC.Jungers.Daafouz2013}
M.~Jungers and J.~Daafouz, ``Guaranteed cost certification for discrete-time
  linear switched systems with a dwell time,'' \emph{IEEE Trans. Autom.
  Control}, pp. 768--772, 2013.

\bibitem{SCC.Heydari2017}
A.~Heydari, ``Optimal switching with minimum dwell time constraint,'' \emph{J.
  Franklin Inst.}, vol. 354, no.~11, pp. 4498--4518, 2017.

\bibitem{SCC.Wardi.Egerstedt.ea2015}
Y.~Wardi, M.~Egerstedt, and M.~Hale, ``Switched-mode systems: gradient-descent
  algorithms with {A}rmijo step sizes,'' \emph{Discrete Event Dyn. Syst.},
  vol.~25, no.~4, pp. 571--599, 2015.

\bibitem{SCC.Bengea.DeCarlo2005}
S.~C. Bengea and R.~A. DeCarlo, ``Optimal control of switching systems,''
  \emph{Automatica}, vol.~41, no.~1, pp. 11--27, 2005.

\bibitem{SCC.Berkovitz1974}
L.~Berkovitz, \emph{Optimal control theory}.\hskip 1em plus 0.5em minus
  0.4em\relax Springer-Verlag New York, 1974.

\bibitem{SCC.Ge1996}
X.~Ge and A.~Nerode, ``Effective content of the calculus of variations i:
  semi-continuity and the chattering lemma,'' \emph{Ann. Pure Appl. Logic,},
  vol.~78, no. 1-3, pp. 127--146, 1996.

\bibitem{SCC.Zangwil1967}
W.~I. Zangwil, ``The piecewise concave function,'' \emph{Manage. Sci.},
  vol.~13, no.~11, pp. 773--945, Jul. 1967.

\bibitem{SCC.Rao.Benson.ea2010}
A.~V. Rao, D.~A. Benson, C.~L. Darby, M.~A. Patterson, C.~Francolin, and G.~T.
  Huntington, ``Algorithm 902: {GPOPS}, a {MATLAB} software for solving
  multiple-phase optimal control problems using the {G}auss pseudospectral
  method,'' \emph{ACM Trans. Math. Softw.}, vol.~37, no.~2, pp. 1--39, 2010.

\end{thebibliography}

\end{document}